\documentclass{article}
\usepackage{spconf,amsmath,graphicx,multirow,amssymb,tablefootnote,hyperref}

\def\x{{\mathbf x}}
\def\c{{\mathbf c}}
\def\L{{\cal L}}

\title{Universal MelGAN: A Robust Neural Vocoder for High-Fidelity Waveform Generation in Multiple Domains}

\name{Won Jang$^1$, Dan Lim$^2$, Jaesam Yoon$^1$}
\address{$^1$Kakao Enterprise Corp., Seongnam, Korea\\
$^2$Kakao Corp., Seongnam, Korea}

\begin{document}
\ninept
\maketitle
\begin{abstract}
We propose Universal MelGAN, a vocoder that synthesizes high-fidelity speech in multiple domains. To preserve sound quality when the MelGAN-based structure is trained with a dataset of hundreds of speakers, we added multi-resolution spectrogram discriminators to sharpen the spectral resolution of the generated waveforms. This enables the model to generate realistic waveforms of multi-speakers, by alleviating the over-smoothing problem in the high frequency band of the large footprint model. Our structure generates signals close to ground-truth data without reducing the inference speed, by discriminating the waveform and spectrogram during training. The model achieved the best mean opinion score (MOS) in most scenarios using ground-truth mel-spectrogram as an input. Especially, it showed superior performance in unseen domains with regard of speaker, emotion, and language. Moreover, in a multi-speaker text-to-speech scenario using mel-spectrogram generated by a transformer model, it synthesized high-fidelity speech of 4.22 MOS. These results, achieved without external domain information, highlight the potential of the proposed model as a universal vocoder.
\end{abstract}
\section{Introduction}
\label{sec:intro}

Vocoders were originally used for speech compression in the field of communication. Recently, vocoders have been utilized in various fields such as text-to-speech\cite{shen2018natural, ren2020fastspeech, lancucki2020fastpitch, lim2020jdi} and voice conversion\cite{liu2018wavenet} or speech-to-speech translation\cite{jia2019direct}. Neural vocoders generate human-like voices using neural networks, instead of using traditional methods that contain audible artifacts \cite{griffin1984signal, kawahara1999restructuring, morise2016world}.

Recently, it has been demonstrated that vocoders exhibit superior performances in generation speed and audio fidelity when trained with single speaker utterances. However, some models face difficulty when generating natural sounds in multiple domains such as speakers, language, or expressive utterances. The ability of these models can be evaluated by the sound quality when the model is trained on data of multiple speakers and the sound quality of the unseen domain (from an out-of-domain source). A vocoder that can generate high-fidelity audio in various domains,  regardless of whether the input has been encountered during training or has come from an out-of-domain source, is usually called a \textit{universal vocoder}\cite{lorenzo2018towards, hsu2019towards}.

MelGAN\cite{kumar2019melgan} is a vocoder based on generative adversarial networks (GANs)\cite{goodfellow2014generative}. It is a lightweight and robust model for unseen speakers but yields lower fidelity than popularly employed models \cite{oord2016wavenet, kalchbrenner2018efficient, prenger2019waveglow}. MelGAN alleviates the metallic sound that occurs mainly in unvoiced and breathy speech segments through multi-scale discriminators that receive different scale waveforms as inputs. However, it has not been implemented efficiently for learning with multiple speakers for a universal vocoder.

In this study, we propose \textit{Universal MelGAN}. The generated waveform of the original MelGAN with audible artifacts appears as an over-smoothing problem with a non-sharp spectrogram. We added \textit{multi-resolution spectrogram discriminators} to the model to address this problem in the frequency domain. Our multi-scale discriminators enable fine-grained spectrogram prediction by discriminating waveforms and spectrograms. In particular, they alleviate the over-smoothing problem in the high frequency band of the large footprint model, enabling the generation of realistic multi-speaker waveforms.

To evaluate the performance of the proposed model, we compare with full-band MelGAN (FB-MelGAN)\cite{yang2020multi} as a baseline and two other vocoders: WaveGlow\cite{prenger2019waveglow} and WaveRNN\cite{lorenzo2018towards}. We designed experiments in both Korean and English for language independency. For evaluation, we prepared multiple speaker utterances that included unseen domain scenarios, such as new speakers, emotions, and languages.

The evaluation results indicate that the proposed model achieved the best mean opinion score (MOS) in most scenarios and efficiently preserved the fidelity in unseen speakers. In addition, the evaluations show that the model efficiently preserves the original speech, even in challenging domains such as expressive utterances and unseen languages. In multi-speaker text-to-speech scenarios, our model can generate high-fidelity waveforms with high MOS, and the model outperforms compared vocoders. This results without any external domain information suggest the possibility of the proposed model as a universal vocoder.

\section{Related work}
\label{sec:prior}
MelGAN\cite{kumar2019melgan} is a GAN-based lightweight vocoder in which the generator comprises transposed convolution layers for upsampling and a stack of residual blocks for effective conversion. Multiple discriminators are trained with different scale waveforms to operate in different ranges. Recently, a modified architecture called FB-MelGAN\cite{yang2020multi} with improved fidelity has been proposed.

WaveGlow\cite{prenger2019waveglow} can directly maximize the likelihood of data based on a normalizing flow. A chain of flows transforms simple distributions (e.g. isotropic Gaussian) into the desired data distribution. It has been shown\cite{maiti2020speaker} that the speaker generalization of WaveGlow obtained high objective scores than other models.

WaveRNN\cite{kalchbrenner2018efficient} is an autoregressive model that generates waveforms using recurrent neural network (RNN) layers. It has been demonstrated that WaveRNN preserves sound quality in unseen domains\cite{lorenzo2018towards}. The robustness of WaveRNN using speaker representations has been discussed\cite{paul2020speaker}.

\begin{figure*}[htb]
\begin{minipage}[b]{0.4\linewidth}
  \centering
  \centerline{\includegraphics[height=6.8cm]{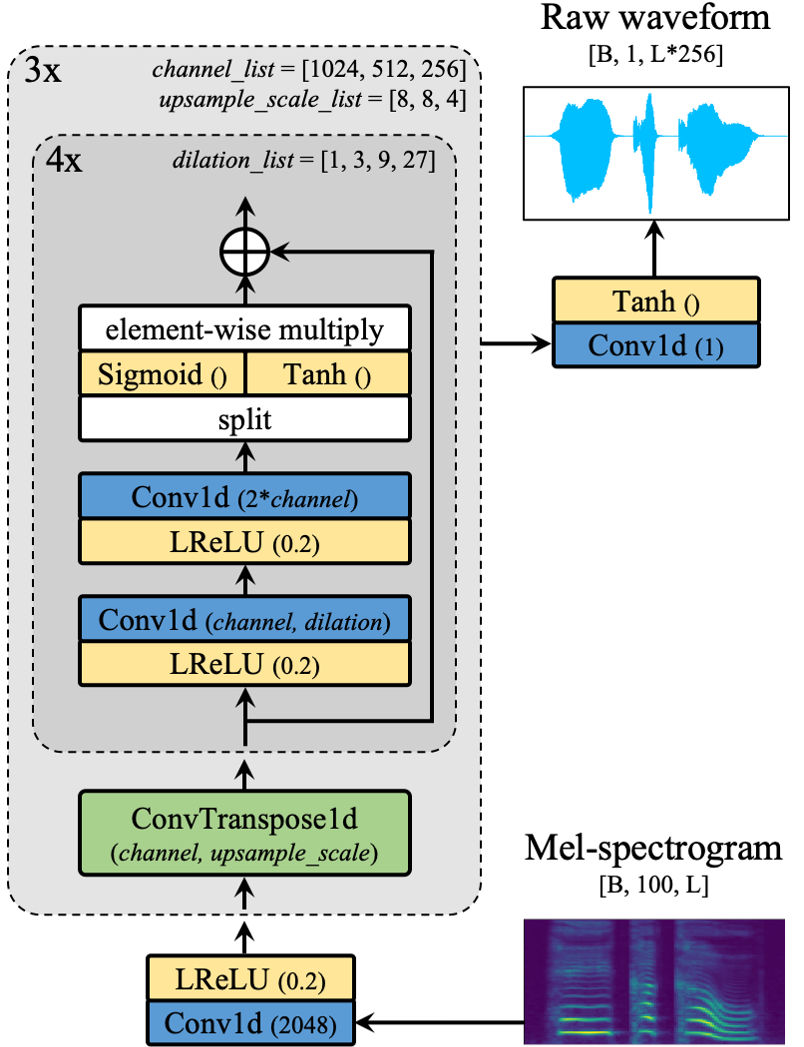}}
  \centerline{(a) Generator}\medskip
\end{minipage}
\begin{minipage}[b]{0.55\linewidth}
  \centering
  \centerline{\includegraphics[height=6.8cm]{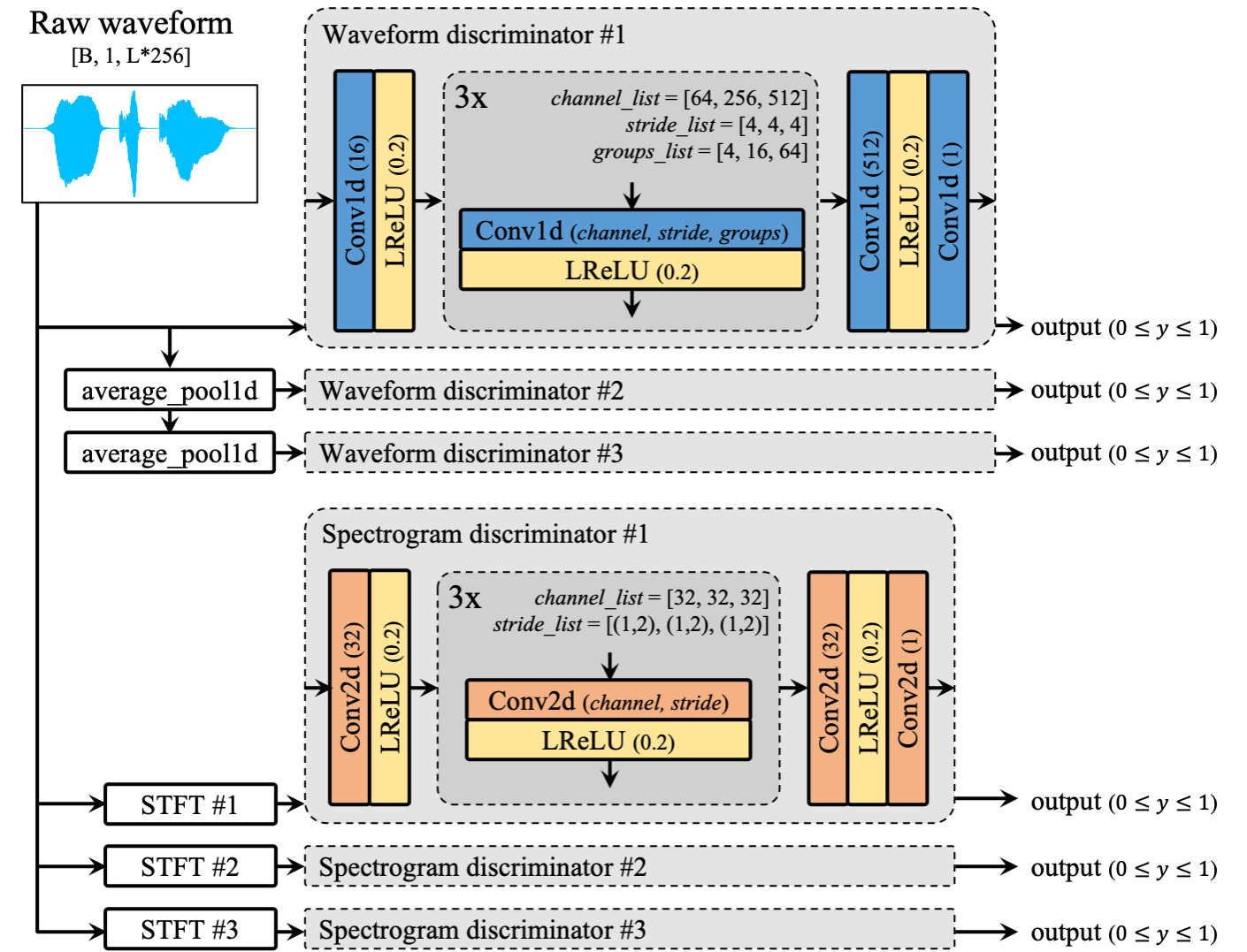}}
  \centerline{(b) Multi-scale waveform and spectrogram discriminators}\medskip
\end{minipage}
\caption{An architecture of the Universal MelGAN.}
\label{fig:model}
\end{figure*}

\section{Description of the proposed model}
\label{sec:proposed}

\subsection{FB-MelGAN}
\label{ssec:fb_melgan}
FB-MelGAN\cite{yang2020multi} has been used as the baseline in this study. The advantages of FB-MelGAN include pre-training of the generator, increased receptive field of residual stacks, and application of multi-resolution short-time Fourier transform (STFT) loss\cite{yamamoto2020parallel} as auxiliary loss. These modifications are effective to achieve better fidelity and training stability.

The multi-resolution STFT loss is the sum of multiple spectrogram losses calculated with different STFT parameter sets. It comprises the spectral convergence loss $\L^{m}_{\mathrm{sc}}(\cdot, \cdot)$ and the log STFT magnitude loss $\L^{m}_{\mathrm{mag}}(\cdot, \cdot)$. These objectives are defined as follows:
\begin{equation}
\label{eq:sc}
\L^{m}_{\mathrm{sc}}(\x, \hat{\x}) =  \frac{\Vert{}\vert{}\mathit{STFT}_{m}(\x)\vert{}-\vert{}\mathit{STFT}_{m}(\hat{\x})\vert{}\Vert{}_{F}}{\Vert{}\vert{}\mathit{STFT}_{m}(\x)\vert{}\Vert{}_{F}}
\end{equation}
\begin{equation}
\label{eq:mag}
\L^{m}_{\mathrm{mag}}(\x, \hat{\x}) = \frac{1}{N}\Vert{}\log{}\vert{}\mathit{STFT}_{m}(\x)\vert{}-\log{}\vert{}\mathit{STFT}_{m}(\hat{\x})\vert{}\Vert{}_{1}
\end{equation}
\begin{equation}
\label{eq:stft}
\L_{\mathrm{aux}}(G)=\frac{1}{M}\sum_{m=1}^M\mathbb{E}_{\x, \hat{\x}}\Big{[}\L^{m}_{\mathrm{sc}}(\x, \hat{\x})+\L^{m}_{\mathrm{mag}}(\x, \hat{\x})\Big{]}
\end{equation}
where $\Vert\cdot\Vert_{F}$ and $\Vert\cdot\Vert_{1}$ denote the Frobenius and L1 norms, and $M$ and $N$ denote the number of STFT parameter sets and the number of elements in the STFT magnitude, respectively. $\vert\mathit{STFT}_{m}(\cdot)\vert$ denotes the STFT magnitude of the $m$-th STFT parameter set. The loss is used to minimize distances between the real data $\x$ and predicted data $\hat{\x}=G(\c)$, using the generator $G$. The overall objectives with auxiliary loss are defined as follows:
\begin{equation}
\label{eq:generator}
\L_{\mathrm{G}}(G, D)=\L_{\mathrm{aux}}(G)+\frac{\lambda}{K}\sum_{k=1}^{K}\mathbb{E}_{\hat{\x}}[(D_{k}(\hat{\x})-1)^2]
\end{equation}
\begin{equation}
\label{eq:discriminator}
\L_{\mathrm{D}}(G, D)=\frac{1}{K}\sum_{k=1}^{K}(\mathbb{E}_{\x}[(D_{k}(\x)-1)^2]+\mathbb{E}_{\hat{\x}}[D_{k}(\hat{\x})^2])
\end{equation}
where $\c$ denotes the mel-spectrogram, and $K$ refers to the number of discriminators $D_{k}$. There is a balancing parameter $\lambda$ that optimizes the adversarial and auxiliary loss simultaneously.

\begin{figure*}[htb]
\begin{minipage}[b]{1.0\linewidth}
  \centering
  \centerline{\includegraphics[width=17.5cm]{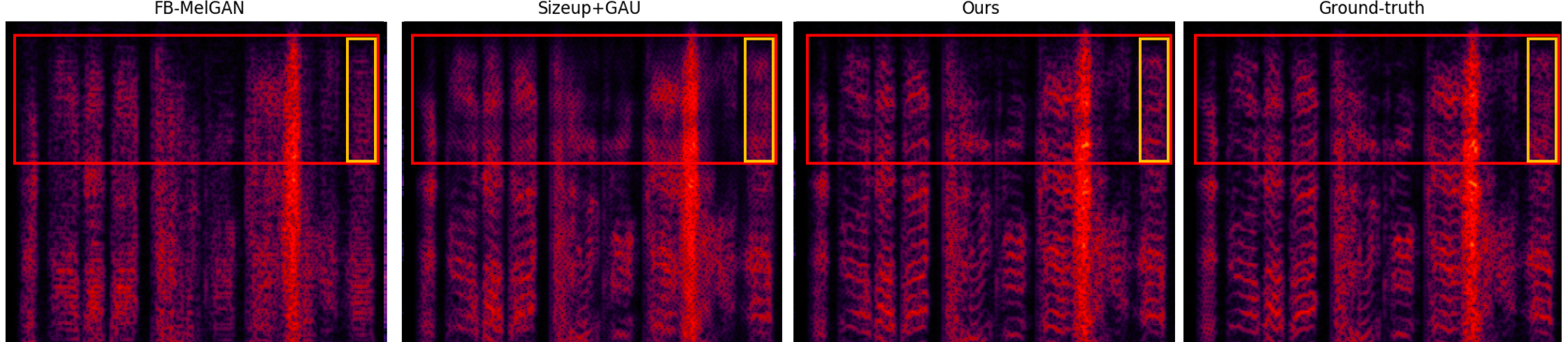}}
  \medskip
\end{minipage}
\caption{Comparison of spectrograms of the generated waveforms in the 6 to 12 kHz bands. (Each waveform has 24 kHz sampling rate.)}
\label{fig:spectrogram}
\end{figure*}

\subsection{Universal MelGAN: Improvements}
\label{ssec:improvements}

We confirmed that FB-MelGAN trained on the utterances of hundreds of speakers generated unsatisfactory sound quality waveforms, unlike when trained on a single speaker's utterances. In FB-MelGAN, multi-scale discriminators alleviate this degradation problem. However, in our multi-speaker experiment, metallic sounds were discernable, especially in unvoiced and breathy segments.

To trade inference speed and quality, we increased the sizes of the hidden channel in the generator by four times and added gated activation units (GAUs)\cite{van2016conditional} to the last layers of each residual stack. Although the expansion of the nonlinearity improved the average quality of the multi-speaker, the over-smoothing problem appeared in the high frequency band, accompanied by audible artifacts.

We assumed that discriminators in the temporal domain may not be sufficient for the problem in the frequency domain. To solve the problem, we propose \textit{multi-resolution spectrogram discriminators} that expand the spectrogram discriminator\cite{su2020hifi}. The objecties for \textit{Universal MelGAN} can be updated as follows:
\vspace{-0.2cm}
\begin{equation}
\label{eq:ours_generator}
\begin{split}
\L^{\prime}_{\mathrm{G}}(G, D)=\L_{\mathrm{aux}}(G)+\frac{\lambda}{K\!+\!M}\big{(}\sum_{k=1}^{K}\mathbb{E}_{\hat{\x}}[(D_{k}(\hat{\x})-1)^2]\\[-3pt]+\sum_{m=1}^{M}\mathbb{E}_{\hat{\x}}[(D^{s}_{m}(\vert\mathit{STFT}_{m}(\hat{\x})\vert)-1)^2]\big{)}
\end{split}
\end{equation}
\vspace{-0.2cm}
\begin{equation}
\label{eq:ours_discriminator}
\begin{split}
\L^{\prime}_{\mathrm{D}}(G, D)=\frac{1}{K\!+\!M}\big{(}\sum_{k=1}^{K}(\mathbb{E}_{\x}[(D_{k}(\x)-1)^2]+\mathbb{E}_{\hat{\x}}[D_{k}(\hat{\x})^2]) \\[-3pt]\!+\!\sum_{m=1}^{M}\!(\mathbb{E}_{\x}[(D^{s}_{m}(\vert\mathit{STFT}_{m}(\x)\vert)\!-\!1)^2]\!+\!\mathbb{E}_{\hat{\x}}[D^{s}_{m}(\vert\mathit{STFT}_{m}(\hat{\x})\vert)^2])\!\big{)}
\end{split}
\end{equation}
where $D^{s}_{m}$ denotes a \textit{spectrogram discriminator} attached to the multi-resolution STFT module. Each $m$-th model uses the previously calculated spectrogram to minimize the $m$-th STFT loss.

Fig. 1. shows the architecture of the Universal MelGAN. The waveforms generated by the large footprint generator in Fig. 1(a) are discriminated by multiple scales for both the waveform and spectrogram in Fig. 1(b).

In Fig. 2, we compared the high frequency bands of the generated waveforms. Although the large foorprint model of the baseline  produces improved resolution, it still faces an over-smoothing problem, especially in high bands above 9 kHz. The proposed model not only alleviates this problem (red) but also generates harmonic shapes that there are not in the ground-truth in some segments (yellow).

\section{Experiments}
\label{sec:experiments}
We designed experiments in both Korean and English. During training, we used reading style and studio-quality internal datasets with 62 speakers and 265k utterances in Korean. We also used the reading style open datasets, LJSpeech-1.1\cite{ito2017lj} and LibriTTS(train-clean-360)\cite{zen2019libritts}, with 905 speakers and 129k utterances in English. To accurately evaluate the robustness, we prepared test sets that included various seen and unseen domains, as shown in Table 1.

To evaluate the seen domain (i.e. speaker), we considered two scenarios: a speaker with relatively more utterances and a speaker with relatively fewer utterances in the total training set. To evaluate whether the generated waveforms of the speaker trained with fewer utterances also preserved the sound quality well, the training set comprised a relatively large set of single speakers (K1 for Korean, E1 for English) and a multi-speaker small set for each speaker (K2 for Korean, E2 for English). Each speaker's test set was used to evaluate the sound quality.

To evaluate the unseen domains, three scenarios that were not included in the training data were considered: speaker, emotion, and language. We prepared utterances of unseen speakers to evaluate the robustness of the speaker (K3 for Korean, E3 for English). We included the following variations of emotional utterances, such as happiness, sadness, anger, fear, disgust, or sports casting in the evaluation sets (K4 for Korean, E4 for English). Ten unseen languages were used to evaluate the language robustness (M1 for both languages). We evaluated the universality of each model using the utterances of these unseen domains.

\begin{table}[ht]
\caption{Datasets for evaluation.}
\label{tab:evaluation_set}
\centering
\begin{tabular}{l|l|l}
\hline
\multicolumn{3}{c}{Korean datasets}                   \\
\hline
Index & Name & Remarks     \\ \hline
K1    & Internal dataset \#1    & Seen: single speaker \\ 
K2    & Internal dataset \#2    & Seen: multi-speaker \\ 
K3    & Internal dataset \#3    & Unseen: speaker \\
\multirow{2}{*}{K4}    & Internal dataset \#4    & \multirow{2}{*}{Unseen: emotion} \\
    & \,+AICompanion-Emotion\tablefootnote{Available at: \url{ http://aicompanion.or.kr/nanum/tech/data_introduce.php?idx=45}} & \\ \hline
\multicolumn{3}{c}{English datasets}        \\
\hline
Index & Name & Remarks     \\ \hline
E1    & LJSpeech-1.1\cite{ito2017lj}    & Seen: single speaker \\ 
E2    & LibriTTS(train-clean-360)\cite{zen2019libritts} & Seen: multi-speaker \\ 
E3    & LibriTTS(test-clean)\cite{zen2019libritts}    & Unseen: speaker \\
E4    & BlizzardChallenge2013\cite{king2013blizzard}    & Unseen: emotion \\ \hline
\multicolumn{3}{c}{Multiple language dataset}  \\
\hline
Index & Name & Remarks     \\ \hline
M1    & CSS10\cite{park2019css10} & Unseen: language   \\ \hline
\end{tabular}
\end{table}

\subsection{Data configurations}
\label{ssec:data_configurations}

All speech samples were resampled to a rate of 24 kHz. We used a high-pass filter at 50 Hz and normalized the loudness to -23 LUFS. 100-band log-mel-spectrograms with 0 to 12 kHz frequency bands were extracted by using a 1024-point Fourier transform, 256 sample frame shift, and 1024 sample frame length. All spectrograms were normalized utterance-wise to have an average of 0 and variance of 1.

\subsection{Model settings}
\label{ssec:settings}

The settings that are not specified follow the original papers or implementations in the footnotes.

When training WaveGlow\cite{prenger2019waveglow}\footnote{\url{https://github.com/NVIDIA/waveglow}} and WaveRNN\cite{lorenzo2018towards}\footnote{\url{https://github.com/bshall/UniversalVocoding}}, GitHub implementations were used for reproducibility. In WaveGlow, we trained up to 1M steps with all settings same as those in the implementation. For inference, a noise sampling parameter $\sigma=0.6$ was used. In WaveRNN, we set up predicting 10-bit $\mu$-law samples. At every 100k step, we reduced the learning rate to half of the initial value of $4e-4$ and trained up to 500k steps.

In FB-MelGAN\cite{yang2020multi}, each upsampling rate was set to 8, 8, and 4 to match the hopping size of 256. In our dataset, stable training was effectively realized by reducing the initial learning rate of the discriminators to $5e-5$. The batch size was set to 48. We trained up to 700k steps using the Adam optimizer.

In Universal MelGAN, the channel sizes of the input layers and the layers inside the three residual stacks in the generator were increased by four times; therefore, they were changed to 2048, 1024, 512, and 256, respectively. GAU was added inside each residual stack, and the channel size of the previous layer was doubled to maintain the same output size after GAU processing. Each spectrogram discriminator has a structure similar to the waveform discriminator (same as the discriminator used in FB-MelGAN), as shown in Fig. 1(b). 1-d convolutions were replaced with 2-d, and all layers have a channel size of 32, 1 group, and 1 dilation. The last two layers have a kernel size of 3, and all other layers have kernel sizes of 9. During operation, the length of the temporal domain was reduced to a stride of 2 over 3 times. Leaky ReLU with $\alpha=0.2$ was used for each activation. The balancing parameter $\lambda$ for all discriminators was set to 2.5. We trained up to 700k steps with the same learning rate, training strategy, batch size, and optimizer as in FB-MelGAN.

\section{Results}
\label{sec:results}
We implemented MOS assessments in which listeners scored naturalness from 1 (negative) to 5 (positive) for each sample. Ten randomly sampled utterances were prepared for each scenario and model, and 150 scores were collected to calculate the mean and 95\% confidence intervals. Fifteen native listeners participated in the Korean evaluation. For the English evaluation, we used a crowed-sourced evaluation via Amazon Mechanical Turk, with more than 15 workers from the US for each scenario.
\let\thefootnote\relax\footnote{Audio samples are available at the following URL: \\ \url{https://kallavinka8045.github.io/icassp2021/}}

\subsection{Seen speakers}
\label{ssec:seen_speakers}

These scenarios consist of evaluation sets for a single speaker with a relatively large training set for each model and for multiple speakers using a relatively small dataset.

Table 2 shows that the proposed model scores higher than most models, and the difference between the two scenarios is the smallest. This is the first result that represents the robustness of the proposed model that generates high-fidelity speech, regardless of whether the speaker's utterances were used frequently during training.

\begin{table}[t]
\centering
\caption{MOS results of each model for seen speakers.}
\label{tab:seen_speakers}
\begin{tabular}{l|cc}
\hline
\multicolumn{3}{c}{Trained with Korean utterances}                   \\ \hline
Model      & Single speaker(K1) & Multi-speaker(K2) \\ \hline
WaveGlow    & 3.42$\pm$0.08  & 3.10$\pm$0.07  \\
WaveRNN     & 3.97$\pm$0.08  & 3.42$\pm$0.08  \\
FB-MelGAN     & 3.25$\pm$0.09  & 2.72$\pm$0.09   \\
Ours         & \textbf{4.19$\pm$0.09}  & \textbf{4.05$\pm$0.08}    \\ \hline
Recordings & 4.33$\pm$0.08  & 4.23$\pm$0.08     \\ \hline
\multicolumn{3}{c}{Trained with English utterances}                  \\ \hline
Model      & Single speaker(E1) & Multi-speaker(E2) \\ \hline
WaveGlow    & 3.65$\pm$0.15  & 3.27$\pm$0.19  \\
WaveRNN     & \textbf{3.85$\pm$0.14}  & 3.70$\pm$0.15  \\
FB-MelGAN   & 3.62$\pm$0.16  & 3.37$\pm$0.17  \\
Ours        & 3.81$\pm$0.15  & \textbf{3.71$\pm$0.15}   \\ \hline
Recordings & 3.89$\pm$0.16  & 3.79$\pm$0.16     \\ \hline
\end{tabular}
\end{table}

\subsection{Unseen domains: speaker, emotion, language}
\label{ssec:unseen_domains}

These scenarios comprise evaluation sets of utterances from domains that were never used for training.

Table 3 shows that the results of our model are the closest to the recordings in most scenarios. Note that most models are not efficient in maintaining the performance in the emotion and language sets, compared to the score of speaker set. However, our model maintains a relatively small score difference. This result represents that our model can preserve sound quality in various unseen domains.

\begin{table}[t]
\centering
\caption{MOS results of each model for unseen domains.}
\label{tab:ood_scenarios}
\begin{tabular}{l|ccc}
\hline
\multicolumn{4}{c}{Trained with Korean utterances}                                                \\ \hline
Model      & Speaker(K3) & Emotion(K4) & Language(M1) \\ \hline
WaveGlow    & 3.27$\pm$0.07 & 3.11$\pm$0.06            & 3.27$\pm$0.07            \\
WaveRNN     & 3.83$\pm$0.08 & 2.46$\pm$0.09            & 2.85$\pm$0.07            \\
FB-MelGAN   & 2.90$\pm$0.08 & 2.60$\pm$0.09            & 2.71$\pm$0.08            \\
Ours        & \textbf{4.15$\pm$0.08} & \textbf{3.91$\pm$0.08}            & \textbf{3.67$\pm$0.07}            \\ \hline
Recordings  & 4.32$\pm$0.08 & 4.31$\pm$0.07            & 3.99$\pm$0.07            \\ \hline
\multicolumn{4}{c}{Trained with English utterances}                                               \\ \hline
Model      & Speaker(E3) & Emotion(E4) & Language(M1) \\ \hline
WaveGlow    & 3.51$\pm$0.16 & 3.13$\pm$0.18            & 3.35$\pm$0.18            \\
WaveRNN     & \textbf{3.80$\pm$0.15} & 3.54$\pm$0.16            & 3.41$\pm$0.18            \\
FB-MelGAN   & 3.48$\pm$0.17 & 3.11$\pm$0.16            & 3.14$\pm$0.18            \\
Ours        & \textbf{3.80$\pm$0.16} & \textbf{3.76$\pm$0.16}            & \textbf{3.58$\pm$0.18}            \\ \hline
Recordings  & 3.95$\pm$0.15 & 4.01$\pm$0.16            & 3.71$\pm$0.17            \\ \hline
\end{tabular}
\end{table}

\subsection{Multi-speaker text-to-speech}
\label{ssec:multi_tts}

To evaluate this scenario, we trained the JDI-T\cite{lim2020jdi} acoustic model with a pitch and energy predictor\cite{ren2020fastspeech, lancucki2020fastpitch} using a dataset with four speakers (containing K1 and subsets of K2). Each trained vocoder was fine-tuned by 100k steps using a pair of the ground-truth waveforms and the predicted mel-spectrograms. Note that we prepared the predicted mel-spectrograms of JDI-T by using the text, reference duration, ground-truth pitch, and energy.

Table 4 shows that Universal MelGAN achieved a real-time synthesis rate of 0.028 real time factor (RTF), with a higher MOS that outperforms other models. We used an NVIDIA V100 GPU to compare the inference speed of each model. All figures were measured without any hardware optimizations (i.e. mixed precision) or methods for accelerating the inference speed with a decrease in sound quality (i.e. batched sampling\cite{kalchbrenner2018efficient} or multi-band generation strategy\cite{yu2019durian}). This result indicates that the proposed model has the ability to synthesize high-fidelity waveforms from text in real-time.

\begin{table}[t]
\centering
\caption{MOS results of multi-speaker text-to-speech and inference speed of each vocoder.}
\label{tab:multi_TTS}
\begin{tabular}{l|c|c}
\hline
\multicolumn{3}{c}{Trained with Korean utterances} \\ \hline
Model    & TTS(K1+K2) & Inference speed \\ \hline
WaveGlow  & 3.36$\pm$0.06 & 0.058 RTF       \\
WaveRNN   & 3.06$\pm$0.10 & 10.12 RTF      \\
FB-MelGAN & 3.43$\pm$0.09 & 0.003 RTF      \\
Ours      & \textbf{4.22$\pm$0.06} & 0.028 RTF    \\ \hline
\end{tabular}
\end{table}

\section{Conclusion}
\label{sec:conclusion}

In this study, we propose Universal MelGAN, a robust neural vocoder for high-fidelity synthesis in multiple domains. We solved the over-smoothing problem that causes a metallic sound, by attaching multi-resolution spectrogram discriminators to the model. Our model is stable while generating waveforms with fine-grained spectrograms in large footprint models. The evaluation results indicate that the proposed model achieved the highest MOS in most seen and unseen domain scenarios. The result demonstrates the universality of the proposed model. For more general use of the model, we will study a lightweight model in the future and apply the multi-band strategy to reduce the complexity while preserving the sound quality.




\bibliographystyle{IEEEbib}
\bibliography{refs}

\end{document}